\documentclass[journal]{IEEEtran}

\usepackage{amsmath}
\usepackage{hyperref}
\usepackage{mathtools, mathrsfs}
\usepackage{caption}
\usepackage{subcaption}
\usepackage{color}
\usepackage{graphicx}
\usepackage{xcolor}

\usepackage{cite}

\begin{document}

\title{Deep Learned Optical Multiplexing for Multi-Focal Plane Microscopy}

\author{Yi Fei Cheng,
        Ziad Sabry,
        Megan Strachan,
        Skyler Cornell, \\
        Jake Chanenson,
        Eva-Maria S. Collins, 
        and Vidya Ganapati
\thanks{Y. F. Cheng, M. Strachan, S. Cornell, J.  Chanenson and V. Ganapati are with the Engineering Department, Swarthmore College, Swarthmore, PA 19081, USA.}
\thanks{Z. Sabry and E.-M. S. Collins are with the Biology Department, Swarthmore College, Swarthmore, PA 19081, USA.}
\thanks{E-mail: vganapa1@swarthmore.edu.}}

\maketitle

\begin{abstract}

To obtain microscope images at multiple focal planes, the distance between the objective and sample can be mechanically adjusted. Images are acquired sequentially at each axial distance. Digital refocusing with a light-emitting diode (LED) array microscope allows elimination of this mechanical movement. In an LED array microscope, the light source of a conventional widefield microscope is replaced with a 2-dimensional LED matrix. A stack of images is acquired from the LED array microscope by sequentially illuminating each LED and capturing an image. Previous work has shown that we can achieve digital refocusing by post-processing this LED image stack. Though mechanical scanning is eliminated, digital refocusing with an LED array microscope has low temporal resolution due to the acquisition of multiple images.

In this work, we propose a new paradigm for multi-focal plane microscopy for live imaging, utilizing an LED array microscope and deep learning. In our deep learning approach, we look for a single LED illumination pattern that allows the information from multiple focal planes to be multiplexed into a single image. We jointly optimize this LED illumination pattern with the parameters of a post-processing deep neural network, using a training set of LED image stacks from fixed, not live, \textit{Dugesia japonica} planarians. Once training is complete, we obtain multiple focal planes by inputting a single multiplexed LED image into the trained post-processing deep neural network. We demonstrate live imaging of a \textit{D. japonica} planarian at 5 focal planes with our method.

\end{abstract}

\section{Introduction and Background}

When looking at a thick specimen through a microscope, parts of the specimen not in the focal plane appear blurred. The blur is increased in parts of the specimen further away from the focal plane. If the axial distance between the specimen and the microscope objective is adjusted, a different axial plane will come into focus. Imaging at multiple focal planes thus allows us to obtain an understanding of the specimen's 3-dimensional structure. Generally, to refocus to a different axial plane, mechanical adjustment of the distance between the sample and optics is necessary. Often it is expensive and time-intensive to mechanically scan the sample. Scanning and collecting images at multiple focal planes is especially difficult for live imaging, as the sample might move during acquisition of an image stack. 

Digital refocusing refers to techniques that emulate axial mechanical scanning without moving the sample or optics. One method to achieve digital refocusing is light-field microscopy \cite{levoy_light_2006}. In light-field microscopy, a lenslet array is used to capture angular information at each spatial point. From the angular and spatial information collected, we can computationally post-process and determine what the image looks like at different focal planes. However, the collection of angular information comes with a tradeoff; we lose spatial resolution.

Another way to capture multiple planes in a single-shot is to use a carefully engineered diffractive optical element \cite{blanchard_simultaneous_1999, abrahamsson_multifocus_2016, he_computational_2018}. With this approach, the pixels of the camera are split into blocks, and each block contains an image from a different focal plane. Each focal plane image has a reduced field-of-view since the image sensor pixels are split up. This method also has the disadvantage of requiring a costly diffractive optical element.

Phase retrieval techniques also capture information that allows for refocusing to multiple planes. Many phase retrieval techniques use interferometry, which requires well-controlled alignment of a laser. Non-interferometric phase retrieval techniques generally require the acquisition of multiple images, sometimes with mechanical scanning \cite{popescu_quantitative_2011}.

Multi-focal plane microscopy is also possible with a light-emitting diode (LED) array microscope. In an LED array microscope, the light source of a conventional widefield microscope is replaced with a 2-dimensional LED matrix. A stack of images can be acquired from the LED array microscope by sequentially illuminating each LED and acquiring an image. The LEDs are far enough away from the sample that single LED illumination can be approximated as a plane wave at a certain angle. From the stack of collected images, different focal plane images can be reconstructed \cite{zheng_microscopy_2011, tian_3d_2014}. The reconstruction of the focal stack is similar to light-field processing, except without the spatial resolution tradeoff. Though mechanical scanning is eliminated, digital refocusing with an LED array microscope still has low temporal resolution, due to the necessity of acquiring one image per LED.

An LED array microscope has proven useful for techniques other than digital refocusing, such as high space-bandwidth reconstruction of complex objects \cite{zheng_wide-field_2013}. For a thin sample, we can use an LED array microscope to collect the image stack (1 image per LED) and then computationally reconstruct both phase and amplitude. In the computational reconstruction, we achieve resolution above the numerical aperture of the microscope objective. The field-of-view is unchanged, but resolution is improved, so the reconstruction has a high space-bandwidth product. However, the acquisition time is long. Multiplexed LED patterns, where multiple LEDs are illuminated at once, allow the number of total images to be reduced \cite{tian_multiplexed_2014, tian_computational_2015}.

To further reduce the number of images needed, deep learning has been applied \cite{robey_optimal_2018, cheng_illumination_2019, kellman_data-driven_2019, kappeler_ptychnet:_2017, nguyen_deep_2018, boominathan_phase_2018, xue_reliable_2019, zhang_fourier_2019}. In our previous work, we demonstrated the use of deep learning to allow object reconstruction with a single multiplexed LED image \cite{robey_optimal_2018, cheng_illumination_2019}. During training, we co-optimized the LED illumination pattern with a post-processing deep neural network. The post-processing deep neural network computationally transforms the single collected image into the complex object we would expect from iterative processing with a full stack of single LED images. By including the LED illumination pattern in the optimization of the neural network, we find that the mutual information between the collected image and the desired object reconstruction increases \cite{robey_optimal_2018}. Our finding highlights the importance of ``physical preprocessing:'' optimizing physical parameters of the imaging system to collect the most informative measurements.

In this work, we apply the idea of physical preprocessing to multi-focal plane microscopy. Our goal is to enable single-shot collection of multiple focal plane images without sacrificing resolution or field-of-view. We achieve this goal by co-optimizing a single multiplexed LED illumination pattern with a post-processing deep neural network. In this approach, our LED illumination pattern and neural network parameters are tuned for a specific sample type.

We make several new contributions in this work. First, we demonstrate single-shot multi-focal plane microscopy without loss of resolution or field-of-view, with minimal hardware modifications to a conventional widefield microscope. Second, we show that the mutual information between the collected measurement and desired multi-focal image reconstruction is increased with our optimized LED illumination pattern. We show this increase in mutual information for real data, as opposed to simulated data in \cite{robey_optimal_2018}. Finally, we demonstrate single-shot live sample imaging at 5 focal planes.

\section{Methods}

\subsection{LED Array Microscope} 

\begin{figure}[htbp]
    \centering
    \includegraphics[scale=0.3]{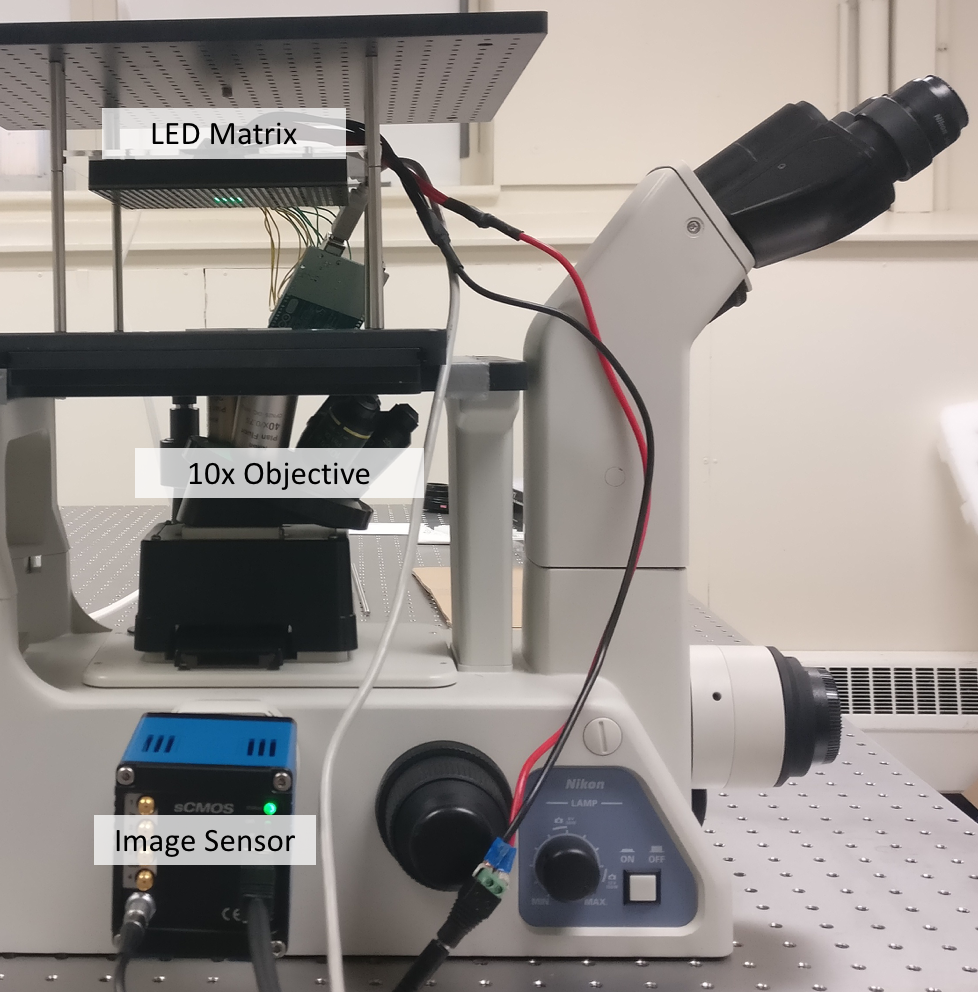}
    \caption{In our optical setup, we replace a conventional microscope light source with a 2-dimensional light-emitting diode (LED) matrix. Each LED has 8 brightness levels, from no light to maximum brightness. Multiple LEDs can be illuminated at once to create different multiplexed illumination patterns.}
    \label{fig:microscope}
\end{figure}

Our experimental setup is shown in Fig.~\ref{fig:microscope}. A programmable matrix of 32 $\times$ 32 LEDs with a 4 mm pitch (Adafruit) is the illumination source of a widefield microscope (Nikon Eclipse TE300). In this work, only the 69 centermost LEDs of the matrix are utilized. The LED matrix is at an axial distance of 69.5 mm away from the sample, and the LED wavelength is centered around 518 nm. Images are collected using a 16 bit image sensor with 2048 $\times$ 2048 pixels, with pixel size of 6.5 $\times$ 6.5 $\mu$m (pco.edge 4.2 LT). The microscope objective has 10$\times$ magnification and numerical aperture of 0.3 (Nikon CFI Plan Fluor).

We assume that the LEDs are far enough away from the sample that the illumination from a single LED approximates a plane wave \cite{zheng_wide-field_2013}. We also assume that the light from the LEDs is mutually incoherent, as in \cite{tian_multiplexed_2014}. To approximate the image from a multiplexed LED illumination pattern, we can take a weighted sum of the single LED image stack:

\begin{equation}
I = \sum \limits_{i=1}^n c_{i} I_i ,
\label{eqn:multiplexed}
\end{equation}

\noindent where $n$ is the total number of illuminated LEDs and $c_i$ is the relative intensity of LED $i$.

\subsection{Digital Refocusing with Shift-Add}

With our LED array microscope, we collect a stack of 69 images. Each image is illuminated using a single LED and captured with an exposure time of 2 seconds. From this stack of images, we can digitally refocus to different focal planes with the shift-add algorithm \cite{zheng_microscopy_2011, tian_3d_2014}. Similar to conventional light-field processing \cite{levoy_light_2006}, the shift-add algorithm makes the assumptions of geometric optics. We assume that the light from a single LED impinging on the sample all comes from a single direction. If the thick 3-dimensional sample is assumed to modulate only the intensity of the incoming light and not the phase, we assume by geometric optics that the light at the image plane remains at the same angle from the optical axis. If we simply add all the LED images together, we get the image at the focal plane. This is the same image as would be obtained by physically illuminating all the LEDs simultaneously. In order to digitally refocus to some axial offset $\Delta z$, we must shear the LED stack and add the resulting images together. The amount each LED image must be shifted depends on both $\Delta z$ and the angle of the light rays. A steeper angle and a greater $\Delta z$ offset both mean a greater shift. 

The lateral position of the LED $\left( x_i, y_i \right) $  determines the angle of illumination $\left( \theta_x, \theta_y \right)$:

\begin{equation} 
 \tan \left( \theta_x \right) =  \frac{x_i}{z},
\end{equation}

\begin{equation} 
\tan \left( \theta_y \right) = \frac{y_i}{z},
\end{equation}

\noindent where $z$ denotes the distance between the sample and the LED matrix. 

To refocus to an axial position $z + \Delta z$, each LED image must be shifted by $\Delta x$ and $\Delta y$ as follows:

\begin{equation} 
\Delta x = \Delta z \tan \left( \theta_x \right),
\end{equation}

\begin{equation} 
\Delta y = \Delta z \tan \left( \theta_y \right).
\end{equation}

To obtain subpixel shifts, we shift in Fourier space as follows: 

\begin{equation} 
I_{\textrm{shift}}= \mathscr{F}^{-1}  \left( \mathscr{F} \left( I \right) e^{j 2 \pi (u\Delta x  + v \Delta y) }  \right),
\end{equation}

\noindent  where $I$ denotes the collected LED intensity image, $\mathscr{F}$ and $\mathscr{F}^{-1}$ denote the Fourier and inverse Fourier transforms respectively, and $u$ and $v$ denote the spatial frequency coordinates. 

To obtain the final refocused image at $z + \Delta z$, we sum all the shifted images:

\begin{equation} 
I_{\Delta z} = \sum_{i=1}^{n} I_{\textrm{shift}, i}
\end{equation}

\noindent  where $n$ denotes the total number of illuminated LEDs and $I_{\textrm{shift}, i}$ is the shifted intensity image from LED $i$.

\begin{figure}[htbp]
    \centering
    \includegraphics[scale=0.22]{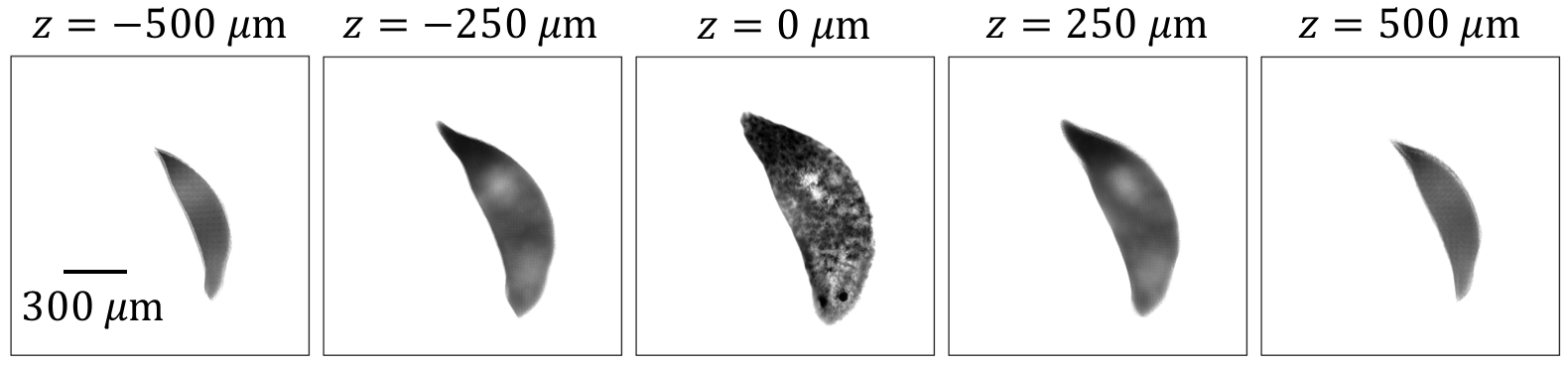}
    \caption{Digital refocusing of a \textit{D. japonica} planarian over 1 mm with application of the shift-add algorithm to the 69 LED image stack.}
    \label{fig:ShiftAdd}
\end{figure}

Fig.~\ref{fig:ShiftAdd} shows an example of the results of the shift-add algorithm. 

\subsection{Deep Learning}

\begin{figure*}[htbp]
    \centering
    \includegraphics[scale=0.35]{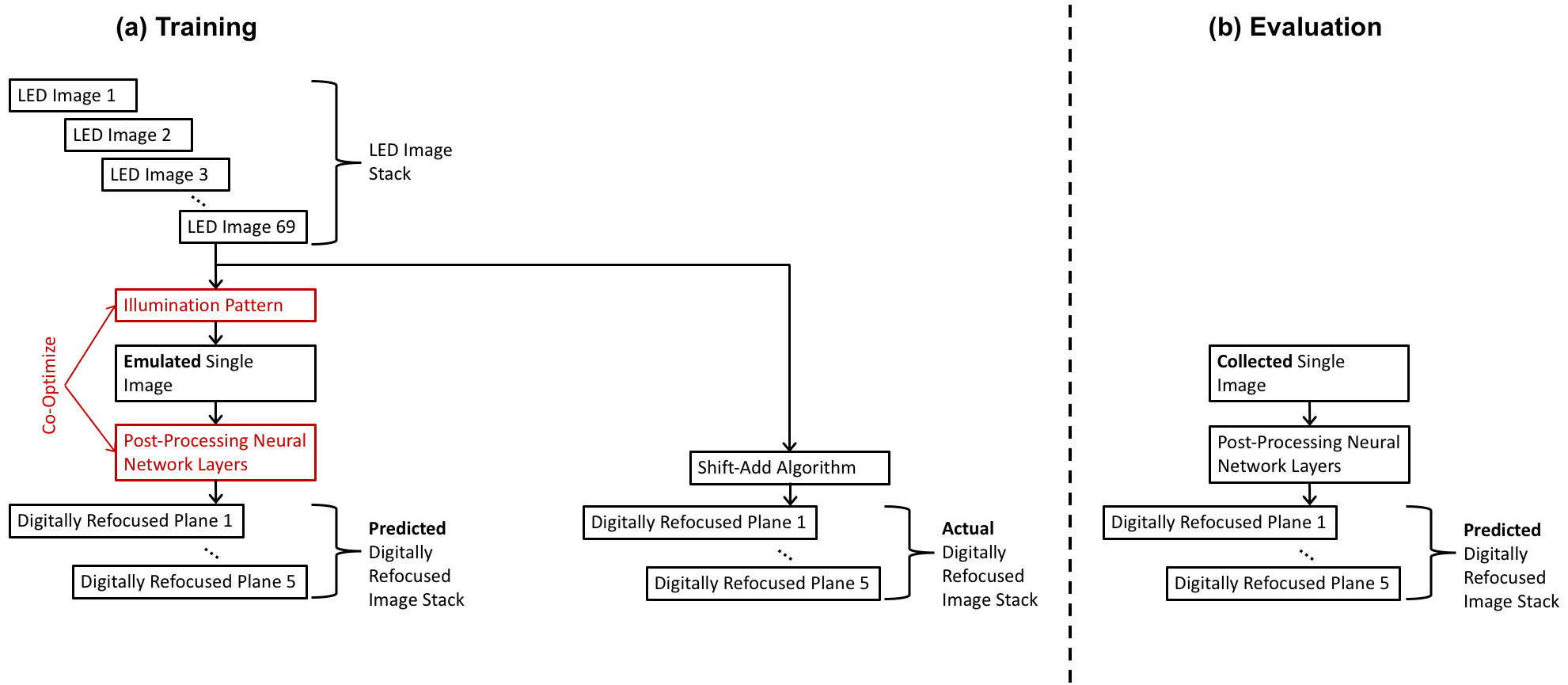}
    \caption{Diagram of training and evaluation of our deep learning framework. (a) In the training step, 69 images are collected for each field-of-view, with each of the 69 images corresponding to illumination with a single LED. During training, we emulate the collection of a single image with a multiplexed illumination pattern. In the illumination pattern, every LED in the matrix is allowed to take a grayscale brightness value. This emulated single image is fed into a post-processing neural network. The output of the post-processing neural network is 5 images, with each image attempting to focus at a different $z$ distance. The illumination pattern and the parameters (weights and biases) of the post-processing neural network are co-optimized during training. The optimization objective is to minimize the difference between the neural network output and a digitally refocused image stack. The digitally refocused image stack of 5 focal planes is calculated by applying the shift-add algorithm to the stack of 69 LED images. (b) In the evaluation step, the optimized illumination pattern is programmed onto the LED matrix of the actual microscope. The single image that is collected from the microscope is then directly fed into the trained post-processing neural network. In the ideal case, the output of the neural network, obtained by a single image, should match the result obtained by collecting 69 separate LED images and applying the shift-add algorithm.}
    \label{fig:TrainingEvaluation}
\end{figure*}

The shift-add algorithm described in the previous section requires a separate image to be collected from each LED. Though we have eliminated mechanical scanning in this method, we still have to collect multiple images. In this work, we take a deep learning approach in order to obtain the same results with just a single collected image. In this deep learning approach, we first collect a training dataset using sequential LED scanning and image acquisition. For each field-of-view in our training dataset, we collect all 69 single LED images. We then construct a neural network graph, where the first layer outputs a linear weighted sum of the 69 input images using Eqn.~\ref{eqn:multiplexed}. The weights of the layer correspond to the relative intensities of each LED in a multiplexed LED illumination pattern. The output of this layer is thus the image you would expect if the multiplexed illumination pattern was used in the LED array microscope. This single image is inputted into a convolutional neural network that outputs $m$ images, where $m$ is the number of focal planes we are trying to reconstruct. The convolutional neural network uses residual connections and is inspired by the network architecture in \cite{mao_image_2016}. The network architecture used in this work is similar to that in \cite{robey_optimal_2018, cheng_illumination_2019}, with the exception that we have $m$ output images, instead of a single complex output.

The LED illumination pattern and convolutional network weights and biases are optimized by training the deep neural network. Our objective in training is to minimize the error between the neural network output and the results of the shift-add algorithm. We calculate the error using a combination of mean-squared error of the images and the first differences, as in \cite{robey_optimal_2018, cheng_illumination_2019}. As in \cite{robey_optimal_2018, cheng_illumination_2019}, in every iteration of training, random Poisson noise is added to the multiplexed LED image, and image saturation is modeled. At the end of training, we normalize the LED weights and modify the exposure time accordingly, so that we obtain the lowest exposure time. As our LED array can only accommodate 8 discrete brightness levels, we also round the LED weights to the nearest level at the end of training.

Our approach is summarized in Fig.~\ref{fig:TrainingEvaluation}.

\subsubsection{Training} 

For the training dataset, we collect LED image stacks in 90 fields-of-view (6 sample slides, 15 field-of-views per sample). Each LED stack consists of 69 images acquired with sequential LED scanning. 

The biological specimens used in this work are \textit{Dugesia japonica} planarians. For the training dataset, \textit{D. japonica} planarians are fixed following standard procedures in \cite{hagstrom_planarian_2018}. Planarians are placed into 100$\%$ glycerol for mounting on custom made tunnel slides using double-stick tape, as in \cite{talbot_smed-dyna-1_2014}.

The single LED image stacks are used to train the neural network graph to determine the optimal multiplexed LED illumination pattern and parameters of the post-processing convolutional neural network.

\subsubsection{Fine-Tuning and Evaluation} 

In the training of the neural network, we model the formation of a single image under a multiplexed LED illumination pattern. To account for errors in this model, a fine-tuning step is performed, wherein a 69 LED image stack is collected as before. In addition, a multiplexed LED image illuminated with the optimized LED pattern is also collected. The multiplexed LED image is inputted directly into the post-processing convolutional neural network, and the weights and biases are re-optimized. Again, we minimize the error between the output of the neural network and the result of the shift-add algorithm on the full 69 LED image stack. 

After fine-tuning is complete, we evaluate the performance of the neural network using fields-of-view that were not previously used in training or fine-tuning. To prevent overfitting, no training or fine-tuning was performed after the evaluation results were obtained.

For the fine-tuning and evaluation steps, we collect images in 40 total fields-of-view (2 sample slides, 20 fields-of-view per sample). The evaluation dataset consists of 8 random field-of-views selected from this dataset.

\begin{figure}[htbp]
    \centering
    \includegraphics[scale=0.27]{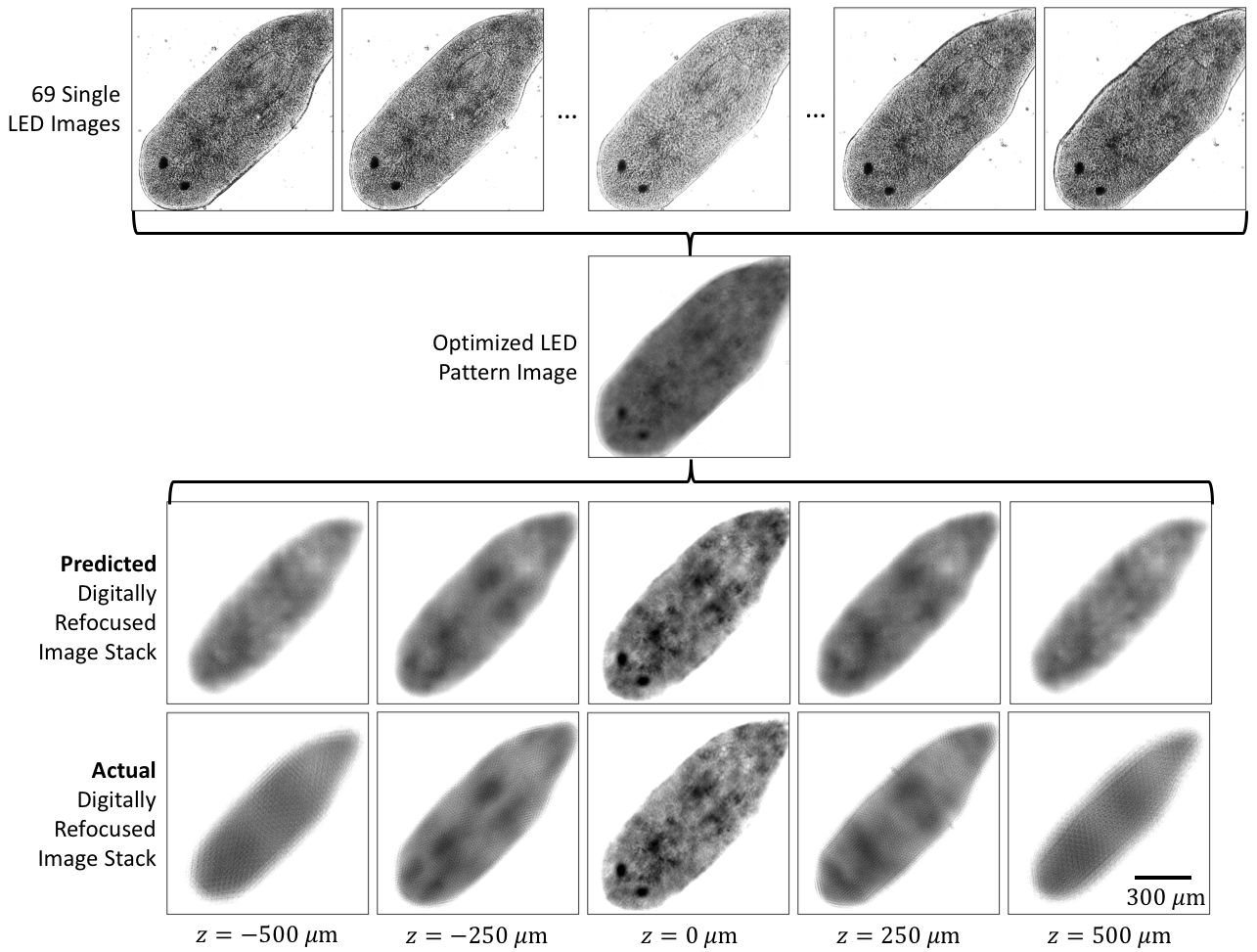}
    \caption{This figure illustrates an evaluation example, a field-of-view that was not used in training or fine-tuning. The top row shows 5 of the total 69 LED images, each illuminated by a single LED of the LED matrix. The next row shows the single LED image obtained by illuminating with the optimized illumination pattern. This optimized LED pattern image is transformed by the post-processing neural network to generate the focal stack shown in the third row. The forth and final row is the focal stack computed by the shift-add algorithm, using all 69 single LED images. Qualitatively, we see that there is agreement between the predicted digitally refocused image stack from the neural network and the actual digitally refocused stack from the shift-add algorithm. We note that the shift-add algorithm assumes geometric optics and doesn't take into account wave effects. Thus, the coherent wave diffraction patterns present in the single LED images are averaged out in the refocused image stack.}
    \label{fig:Pipeline}
\end{figure}

Fig.~\ref{fig:Pipeline} illustrates an evaluation example at various points in the imaging pipeline.  

\subsubsection{Live Imaging}
After the neural network is trained and fine-tuned, we collect videos of live, unfixed \textit{D. japonica} planarians with the optimized LED illumination pattern. We then feed each frame of the video into the post-processing neural network layers to reconstruct 5 focal plane images per frame. In this step, we no longer need the full 69 LED image stack. By keeping the LED matrix statically illuminated with the optimized LED illumination pattern, we can perform single-shot, live imaging of 5 focal planes without loss in resolution or field-of-view.

For live imaging, planarians are mounted in a drop of 1$\times$ Instant Ocean (IO, Blacksburg, VA, USA) and then most of the water is removed.

\section{Results and Discussion}

\subsection{Comparison of Shift-Add and Deep Learning}

\begin{figure}
    \centering
     \begin{subfigure}{1\textwidth}
        \includegraphics[scale=0.28]{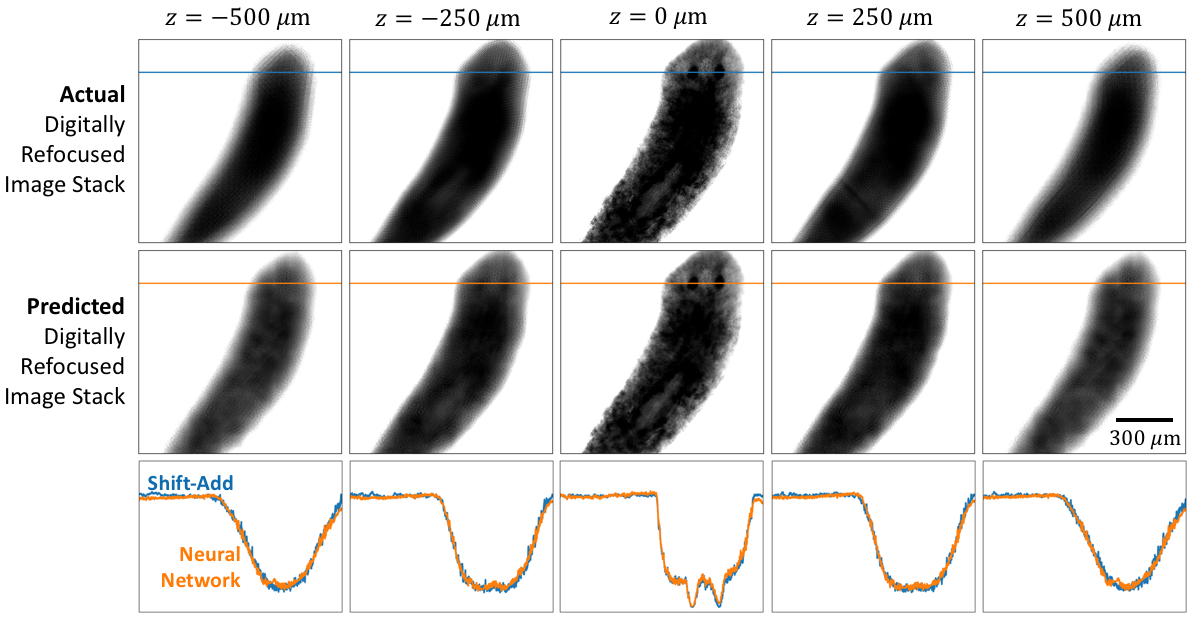}
    \end{subfigure}
    
     \begin{subfigure}{1\textwidth}
        \includegraphics[scale=0.28]{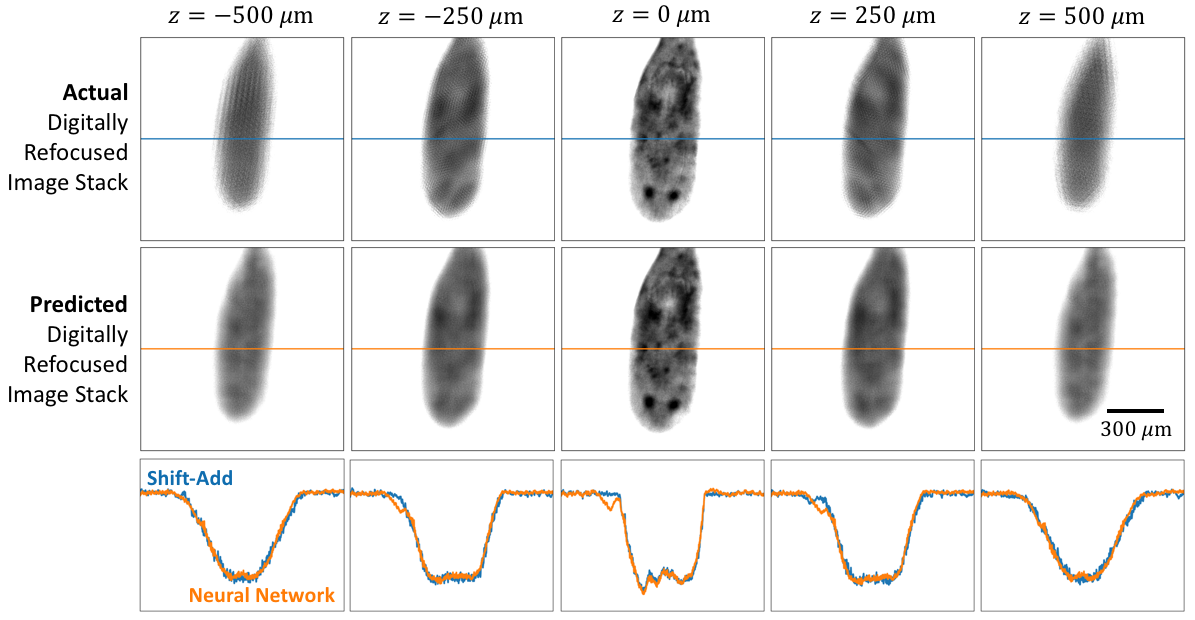}
    \end{subfigure}
 
     \begin{subfigure}{1\textwidth}
        \includegraphics[scale=0.28]{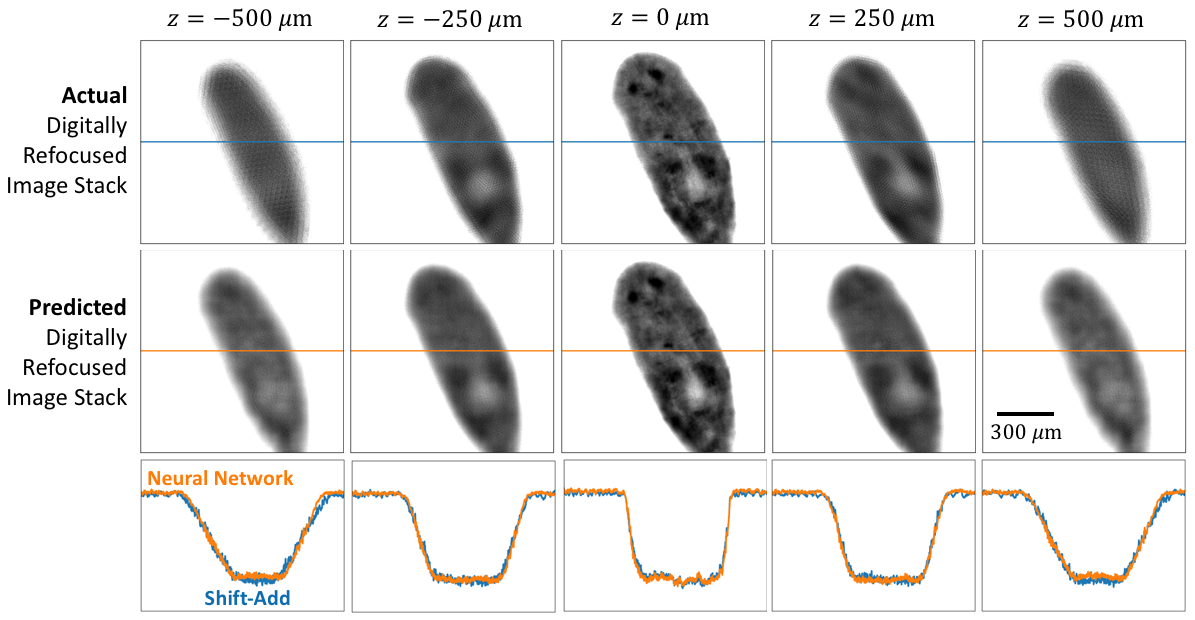}
    \end{subfigure} 
    
    \caption{Comparisons between the actual digitally refocused image stack, using 69 single LED images with the shift-add algorithm, and the predicted digitally refocused stack, using only the single image collected with the optimized LED illumination pattern and neural network post-processing. The linescans show good fidelity of the neural network output to the ground truth shift-add result.}
    \label{fig:Comparison}
\end{figure}

In Fig.~\ref{fig:Comparison} we show a comparison of results between the shift-add algorithm (requiring the full LED stack of 69 images) and the deep learning approach proposed in this work (requiring a single multiplexed LED image) for 3 fields-of-view in the evaluation dataset. The linescans show a good match between the 2 methods. We note that it appears that the neural network approach filters out some random noise present in the shift-add algorithm results. This noise filtering agrees with the results in \cite{lehtinen_noise2noise:_2018}, where the authors demonstrate that a neural network cannot learn zero-mean noise.  

\subsection{Mutual Information}

With a small simulated dataset, we show in \cite{robey_optimal_2018} that optimizing the physical parameters of data collection increases the mutual information between the measurement and the desired final output. We call  data collection with these trained parameters ``optimal physical preprocessing.'' Optimal physical preprocessing allows for less data to be collected without harming the final reconstruction result. This lets us create faster and cheaper imaging systems with no loss in image quality. In this work, we show a mutual information increase with optimal physical preprocessing for a larger dataset of real images. 

We determine the mutual information between 16 $\times$ 16 pixel patches of the multiplexed LED images and the corresponding 5 focal plane stacks of size 5 $\times$ 16 $\times$ 16 pixels. Mutual information quantifies how much the multiplexed LED illumination images ``tell'' us about the 5 focal plane stack. If the mutual information matches the entropy of the focal plane stacks, we should be able to perfectly reconstruct the 5 focal planes from the multiplexed LED image. Mutual information is calculated by: 

\begin{equation}
\sum_{y \in Y} \sum_{x \in X} p \left( x,y \right) \log_2 \left( \frac{p \left( x,y \right)}{ p(x) p(y) } \right), 
\end{equation}

\noindent where $p(x)$ is the marginal probability distribution of 5 focal plane stacks, $p(y)$ is the marginal probability distribution of the multiplexed LED images, and $p(x,y)$ is the joint probability distribution. The Non-Parametric Entropy Estimation Toolbox (NPEET) \cite{steeg_non-parametric_2018} was used to estimate the probability distributions and calculate the mutual information. For every $x \in X$, we generated 10 samples of $y$ with random Poisson noise to approximate the distribution $p(y \mid X = x)$. 

\begin{figure}[htbp]
    \centering
    \includegraphics[scale=0.3]{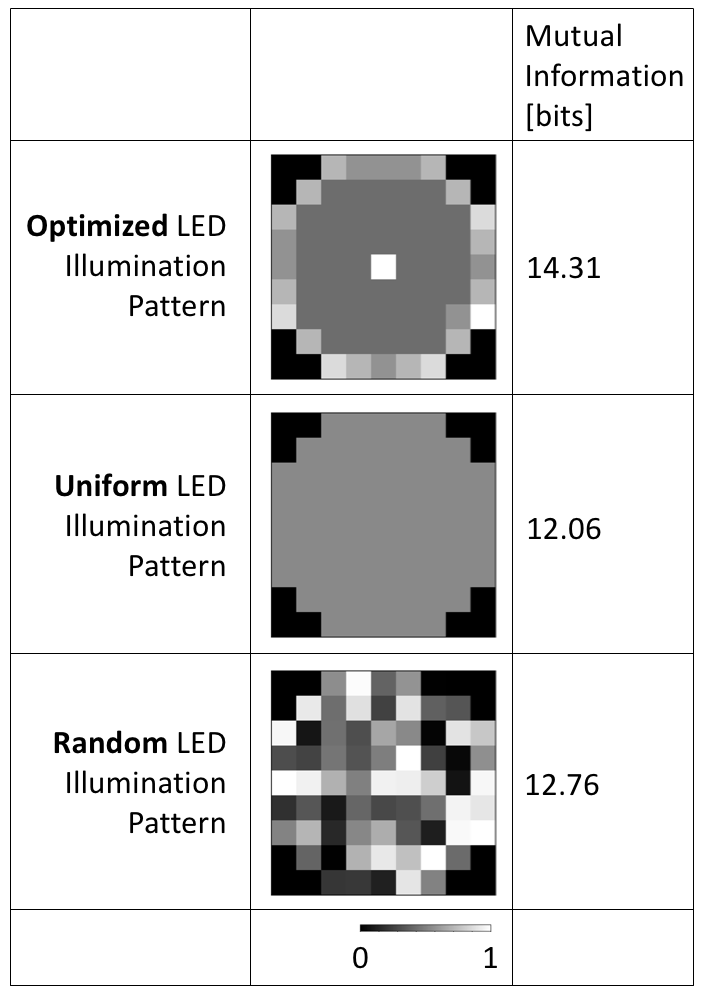}
    \caption{Approximately 300,000 image patches of 16 $\times$ 16 pixels from the evaluation dataset were used to calculate the mutual information of the optimized multiplexed image and the focal stack. The higher the mutual information, the more we know about the focal stack from the optimized optical element image. If the mutual information is the same as the entropy of the focal stack, the optimized optical element image gives us all the information to determine the focal stack exactly. We also calculate the mutual information using a uniform LED illumination pattern and 5 different random LED patterns. The uniform and random LED patterns are normalized such that they have the same average value as the optimized pattern. The mutual information calculated with the optimized LED pattern and the uniform pattern are shown. We show the random LED pattern that achieved the highest mutual information of the 5 random patterns. In this dataset of image patches, the entropy of the focal stack is 17.84 bits. Thus, we cannot expect perfect reconstruction from any of the 3 LED patterns shown, but we can expect the best reconstruction from the optimized LED illumination pattern.}
    \label{fig:MutualInfo}
\end{figure}

We show the mutual information with different multiplexed LED illumination patterns in Fig.~\ref{fig:MutualInfo}, comparing mutual information among optimized, uniform, and random LED illumination patterns. We observe the highest mutual information for the optimized LED illumination pattern. Our observation validates the simulation result in \cite{robey_optimal_2018} with real data and underscores the importance of optimal physical preprocessing. In our joint optimization procedure of the illumination pattern and the post-processing parameters,  the LED illumination pattern changes so that more information about the desired 5 focal plane stack is physically encoded in the single multiplexed LED image.

\subsection{Live Imaging}

\begin{figure*}[htbp]
    \centering
    \includegraphics[scale=0.6]{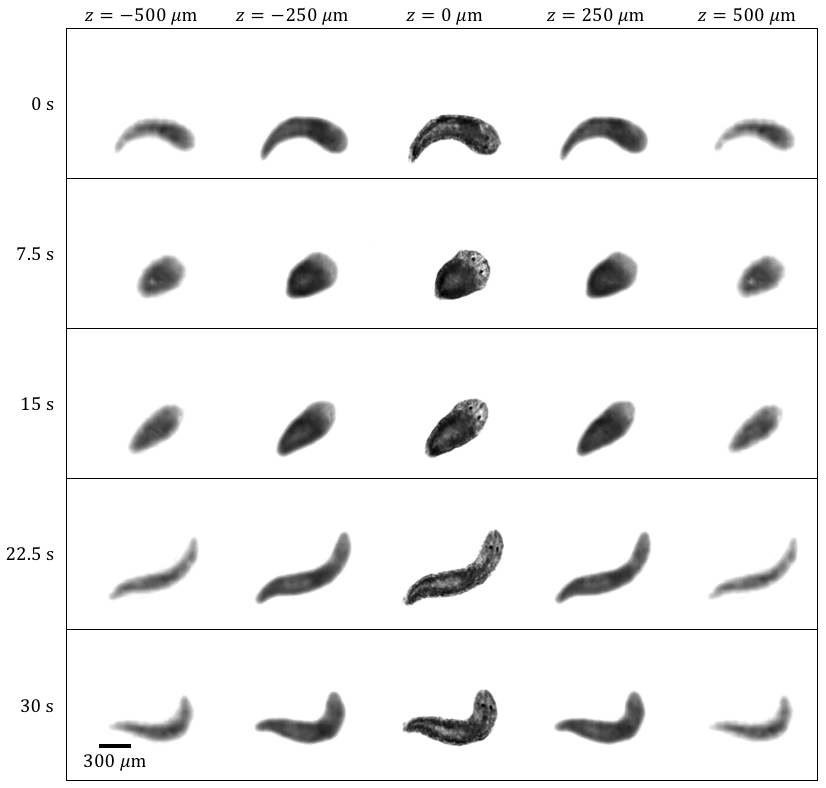}
    \caption{Our deep learning approach allows us to obtain a 5-plane focal stack with single-shot imaging. This allows for live imaging of a \textit{D. japonica} planarian. Time lapse of live video shown, supplementary video available \href{https://youtu.be/7BwI9eK1b1o}{{\color{blue}\underline{online}}}. Each row depicts the planarian at one time point, at 5 different focal planes.}
    \label{fig:TimeLapse}
\end{figure*}

The deep learning approach in this work allows us to train on fixed samples but evaluate on live samples. The training step requires collection of the full 69 LED image stack (1 image per LED) per field-of-view. Once training is complete, only a single image is needed for multiple focal-plane reconstruction. With the trained LED illumination pattern and post-processing neural network parameters, we demonstrate live imaging of a \textit{D. japonica} planarian. Fig.~\ref{fig:TimeLapse} shows time-lapse results with 192 ms exposure, resulting in a rate of 5.2 frames/s. We note that with brighter LEDs we could reduce the exposure time and achieve a higher frame rate.

\section{Conclusions}

In this work, we demonstrate proof-of-concept of a single-shot method to obtain multiple focal stacks without loss in image quality, enabling imaging of live, moving samples. Multiple focal plane imaging allows for understanding of the 3-dimensional structure of the sample. While the proof-of-concept we demonstrate is promising, there are a number of limitations that we plan to address in future work.

One limitation of this work is the use of the shift-add algorithm in generating the ground truth focal stacks. The shift-add algorithm makes the assumptions of geometric optics, leading to artifacts in the refocused planes. The  ground truth could be collected by mechanical axial scanning instead. Another method is to use a phase mask in Fourier space on the imaging side in order to reconstruct the refocused planes using the assumptions of  scalar wave optics \cite{ou_aperture_2016}.

It should also be noted that though 3-dimensional information is available from multiple focal plane stacks, 3-dimensional reconstruction cannot be directly inferred. In fluorescence microscopy, deconvolution algorithms are used to extract point emitter density in 3 dimensions from focal plane stacks \cite{sibarita_deconvolution_2005}. This work considers transmission light microscopy, where the aim is to recover the refractive index at every point in 3-dimensional space. The scattering from a 3-dimensional structure can be approximated with the first Born approximation \cite{horstmeyer_diffraction_2016, ou_aperture_2016, ling_high-throughput_2018, li_high-speed_2019} or the multi-slice model \cite{li_separation_2015, tian_3d_2015}. The deep learning approach outlined in this paper can be applied to directly obtaining the 3-dimensional structure of the sample from a single multiplexed LED image, using one of these approximate forward models to iteratively solve for the ground truth 3-dimensional refractive index.

Another limitation of this work is that we cannot change the number and spacing of the focal planes after the training step. This could be solved if we could reconstruct the entire 69 image stack from a single multiplexed LED image. Then, refocusing to any number of planes and spacings from live imaging data could be done at any time after data collection. We note that we were not able to achieve full mutual information from a single multiplexed LED image for 5 focal planes. Reconstructing the 69 single LED image stack may require 2 or more multiplexed LED images. Future work will determine how many multiplexed images are needed for full mutual information between the data collected and desired reconstruction.

To get single-shot imaging of multiple depth planes without loss of resolution or field-of-view, we trade-off generality of the imaging method. Our method assumes that the object imaged comes from the same probability distribution as the training set of objects. Further work is needed to answer questions such as: how well does this method work, when training on objects from one probability distribution, and evaluating on a different probability distribution? Finally, future work will quantify exactly how many samples are needed for training.

\section*{Acknowledgments}

We acknowledge funding from the Lang Center at Swarthmore College, the Surdna Foundation (Y. F. C.), the Swarthmore College Summer Research Fellowship (M. S.), and NSF CAREER Grant 1555109 (E.-M. S. C.). We thank Leonora Blodgett for helpful comments on the manuscript.

\bibliographystyle{IEEEtran}

\bibliography{paperV2.bib}

\end{document}